# Simultaneously giant enhancement of Förster resonance energy transfer rate and efficiency based on plasmonic excitations

**Jun Ren, Tong Wu, Bing Yang and Xiangdong Zhang***
School of Physics and Beijing Key Laboratory of Nanophotonics & Ultrafine Optoelectronic Systems, Beijing Institute of Technology, 100081, Beijing, China.
* *Corresponding author: zhangxd@bit.edu.cn*

## Abstract

We present a first-principles calculation on the rate and efficiency of Förster resonance energy transfer (FRET) from a donor to an acceptor when they are located in the hotspots of nanoparticle clusters. Nonlocal effect has been considered by using a hydrodynamic model. It is found that FRET rate and efficiency can be enhanced simultaneously by more than 9 and 3 orders of magnitude, respectively. The physical origins for these phenomena have been disclosed. Two opposite phenomena, the energy transfer rate is independent or dependent of the local density of optical states (LDOS), have been observed in the same system under different conditions. These findings not only help us to understand the unresolved debate on how the FRET rate depends on the LDOS, but also provide a new way to realize ultrafast energy transfer process with ultrahigh efficiency.

## I. INTRODUCTION

Resonance energy transfer between molecules or quantum dots has been extensively studied in nanophotonics, which has various important applications ranging from biological detections [1,2] and sensors [3,4] to quantum information science [5-7]. Excitation energy transfer from donor to acceptor molecules separated by a distance $r$ may proceed in three ways, Dexter transfer [8], Förster transfer [9] and radiative transfer [10]. In these ways, Dexter transfer has an effective range $r < 3nm$ and radiative transfer is the simple emission and absorption of a photon, has the longest range and dominates in the $r > \lambda$ range, in which $\lambda$ is the wavelength. Förster resonance energy transfer (FRET) dominates when the distance is much less than the wavelength, which is derived from static dipole-dipole interaction, and is the most important energy transfer channel in the subwavelength range. The FRET plays key role in many processes like photosynthesis [11-13], bioimaging [14,15], lighting [16,17] and photovoltaics [18-20]. Among these processes, the FRET rate and efficiency are the two key parameters. It is very important to obtain simultaneous large FRET rate and efficiency for these applications.

Unfortunately, the FRET rate behaves as $r^{-6}$, when the distance between the two molecules extends to dozens of nanometers, the FRET rate and efficiency are very small. How to improve and control them becomes a key problem in recent years. Many studies focused on the response of photonic environments like photonic crystals [21], resonators [22,23] and plasmonic nanostructures [24-44] on FRET rate and efficiency. However, the contradictory results have been obtained. Some have demonstrated that the FRET rate depends on photonic environments [25,35,45], and some have found it independent of the environments [27,28,37]. As for the FRET efficiency, all these investigations have shown that it cannot be improved with the increase of local density of optical states (LDOS). Simultaneously large enhancement of the FRET rate and efficiency between a single pair of donor and acceptor in the same electromagnetic environment has not been realized so far.

In this work, we study the FRET rate and efficiency from a donor to an acceptor when they are located in the hotspots of nanosphere clusters using a rigorous first-principles electromagnetic Green's tensor technique. We have found that the FRET rate and efficiency can be enhanced simultaneously by several orders of magnitude at suitable parameters. The FRET rate may be related to the LDOS or not, depending on the condition. The rest of this paper is arranged as follows. In Sec. II, we present the theory and method. Simultaneously giant enhancements of FRET rate and efficiency are discussed in Sec. III. The effect of nanoparticle sizes and nonlocal response on FRET rate and efficiency are described in Sec. IV. The relation between the LDOS and the FRET rate is analyzed in Sec. V. A summary is given in Sec. VI.

## II. THEORY AND METHOD

We consider an excited donor molecule D and a ground-state acceptor A located in the gaps of a nanosphere trimer as shown on the left top of Fig. 1, where the position vectors of the two molecules are denoted by $\vec{r}_D$ and $\vec{r}_A$. The right inset on top of Fig. 1 shows the spectral overlap between the normalized donor emission spectral $f_D(\omega)$ and acceptor absorption spectral $f_A(\omega)$ ($\int_0^\infty f_{D(A)}(\omega)\,\mathrm{d}\omega = 1$), $\omega_D$ and $\omega_A$ are central frequencies of the donor and acceptor, respectively, $\omega_{\max}$ represents the frequency of maximum overlap. In present case, the radii of spheres do not exceed tens of nanometer, are much less than the molecular transition wavelength, nonradiative energy transfer namely FRET dominates. Total energy transfer rate between D and A under the electric dipole approximation has the form[46]

$$K_T = \frac{2|\vec{p}_D|^2}{\hbar}\int_0^\infty f_D(\omega)\,\mathrm{Im}[\alpha_A(\omega)]\left(\frac{\omega^2}{\varepsilon_0 c^2}\right)^2 \left|\vec{n}_A\cdot\vec{G}(\vec{r}_A,\vec{r}_D;\omega)\cdot\vec{n}_D\right|^2 \mathrm{d}\omega\,, \tag{1}$$

where $\vec{G}(\vec{r},\vec{r}\,';\omega)$ is the classical Green's tensor of the system, which is the solution of the partial differential equation $\left[\nabla\times\nabla\times - k_0^2\varepsilon(\vec{r})\right]\vec{G}(\vec{r},\vec{r}\,';\omega) = \vec{\mathrm{I}}\,\delta(\vec{r}-\vec{r}\,')$. The $\varepsilon(\vec{r})$ is the dielectric function in position $\vec{r}$, and $\vec{p}_D$ is the dipole moment of the donor molecule, $\vec{\mathrm{I}}$ is the unit tensor. The $\vec{n}_D$ and $\vec{n}_A$ represent directions of dipole moments for the donor and acceptor, respectively. The $\mathrm{Im}[\alpha_A(\omega)]$ is imaginary part of polarizability of the acceptor A ($\mathrm{Im}[\alpha_A(\omega)] = \frac{\pi|\vec{p}_A|^2}{\hbar c^2} f_A(\omega)$). Decay rate of the donor can be expressed with the Green's tensor as[46]

$$K_D(\vec{r}_D) = \frac{2|\vec{p}_D|^2}{\hbar}\int_0^\infty f_D(\omega)\frac{\omega^2}{\varepsilon_0 c^2}\mathrm{Im}\left[\vec{n}_D \cdot \vec{\mathrm{G}}\left(\vec{r}_D,\vec{r}_D;\omega\right)\cdot \vec{n}_D\right]\mathrm{d}\omega\,. \tag{2}$$

The expansion of dynamic Green's tensor contains three terms[37]

$$\vec{G}\left(\vec{r},\vec{r}';\omega\right) = \vec{G}_R\left(\vec{r},\vec{r}';\omega\right) + \vec{G}_S\left(\vec{r},\vec{r}';\omega\right) + \frac{1}{\varepsilon(\vec{r})(\omega/c)^2}\delta(\vec{r}-\vec{r}')\vec{\mathrm{I}}\,, \tag{3}$$

in which the third term can be omitted in the energy transfer process due to $\vec{r} \neq \vec{r}'$. The first term $\vec{G}_R$ corresponds to the radiative dipole-dipole interaction, represents the radiation process, namely the donor located at $\vec{r}$ emits a photon and the acceptor located at $\vec{r}'$ absorbs it. The second term $\vec{G}_S$ corresponds to the static dipole-dipole interaction, represents the nonradiative energy transfer process (namely FRET). Under the quasi-static approximation, the static Green's tensor can be obtained from the dynamical Green's tensor[37]

$$\vec{G}_S = \lim_{\omega\to 0}\frac{\omega^2}{\varepsilon_0 c^2}\vec{G}\,. \tag{4}$$

The detailed calculations of Green's tensor under the quasi-static approximation can be found in supplementary material (see Eq. (A14)). In such a case, the total excitation energy transfer degenerates to the FRET. Then, from Eq. (1), the FRET rate is expressed as

$$K_F = \frac{2|\vec{p}_D|^2}{\hbar}\int_0^\infty W_F\,\mathrm{d}\omega \tag{5}$$

with

$$W_F = f_D(\omega)\,\mathrm{Im}\left[\alpha_A(\omega)\right]\left|\vec{n}_A \cdot \vec{G}_S\left(\vec{r}_A,\vec{r}_D;\omega\right)\cdot \vec{n}_D\right|^2. \tag{6}$$

At the same time, under the quasi-static approximation, the donor decay rate from Eq. (2) turns to

$$K_D(\vec{r}_D) = \frac{2|\vec{p}_D|^2}{\hbar}\int_0^\infty W_D\,\mathrm{d}\omega \tag{7}$$

with

$$W_D = f_D(\omega)\left\{\frac{\omega^3}{6\pi\varepsilon_0 c^3}\sqrt{\varepsilon_M} + \mathrm{Im}\left[\vec{n}_D \cdot \vec{G}_S\left(\vec{r}_D,\vec{r}_D;\omega\right)\cdot \vec{n}_D\right]\right\}, \tag{8}$$

in which $\varepsilon_M$ is the relative dielectric constant of the background medium, in this work the background is taken as water ($\varepsilon_M = 1.77$). From Eqs. (6) and (8), the enhancements of the FRET rate and donor decay rate are written as

$$W_F / W_{F0} = \left|\vec{n}_A \cdot \vec{G}_S\left(\vec{r}_A, \vec{r}_D; \omega\right) \cdot \vec{n}_D\right|^2 / \left|\vec{n}_A \cdot \vec{G}_S^0\left(\vec{r}_A, \vec{r}_D; \omega\right) \cdot \vec{n}_D\right|^2 \quad (9)$$

and

$$W_D / W_{D0} = 1 + \mathrm{Im}\left[\vec{n}_D \cdot \vec{G}_S\left(\vec{r}_D, \vec{r}_D; \omega\right) \cdot \vec{n}_D\right] / \left(\frac{\omega^3}{6\pi\varepsilon_0 c^3}\sqrt{\varepsilon_M}\right). \quad (10)$$

Here $W_{F0}$ and $W_{D0}$ are the corresponding results without nanosphere clusters, and $\vec{G}_S^0\left(\vec{r}_A, \vec{r}_D; \omega\right)$ represents the static Green's tensor in such a case, the detailed calculation for it is given in the supplementary material (see Eq. (A15)). As for $\vec{G}_S\left(\vec{r}_D, \vec{r}_D; \omega\right)$ in Eqs. (8) and (10), it vanishes in the case without nanospheres. The FRET efficiency is defined as[47]

$$\eta_F = (1 + K_D / K_F)^{-1}, \quad (11)$$

which represents the ratio of Förster transfer rate $K_F$ to the total decay rate $(K_D + K_F)$. Based on the above equations, the FRET rate and efficiency in nanosphere clusters can be obtained exactly through numerical calculations.

## III. SIMULTANEOUSLY GIANT ENHANCEMENT OF THE RATE AND EFFICIENCY

Figure 1(a)-(c) show the calculated results of $W_F / W_{F0}$ and $W_D / W_{D0}$ as a function of the frequency for a pair of donor and acceptor located in the gaps of the nanosphere trimer with the separation distances d=1nm, 2nm and 4nm, respectively. The corresponding results for the efficiency enhancement $\eta_F / \eta_{F0}$ as a function of the maximum spectral overlap frequency are given in Fig. 1(d)-(f), here $\eta_{F0}$ is the efficiency without nanospheres. The orientations of electric dipole moments of molecules are set to align with the symmetry axis of the trimer. According to Pustovit and Shahbazyan[36], spectral functions of the donor emission and acceptor absorption are taken as the normalized Lorentzian line-shape with half width 0.08eV centered at $\omega_D = \omega_{\max} - \delta$ and $\omega_A = \omega_{\max} + \delta$, at the present case $\delta = 0.1eV$. The radii of three Ag spheres in the trimer are taken as R=10nm. For the dielectric functions of Ag, the Johnson's data were adopted (the absorption loss is included)[48].

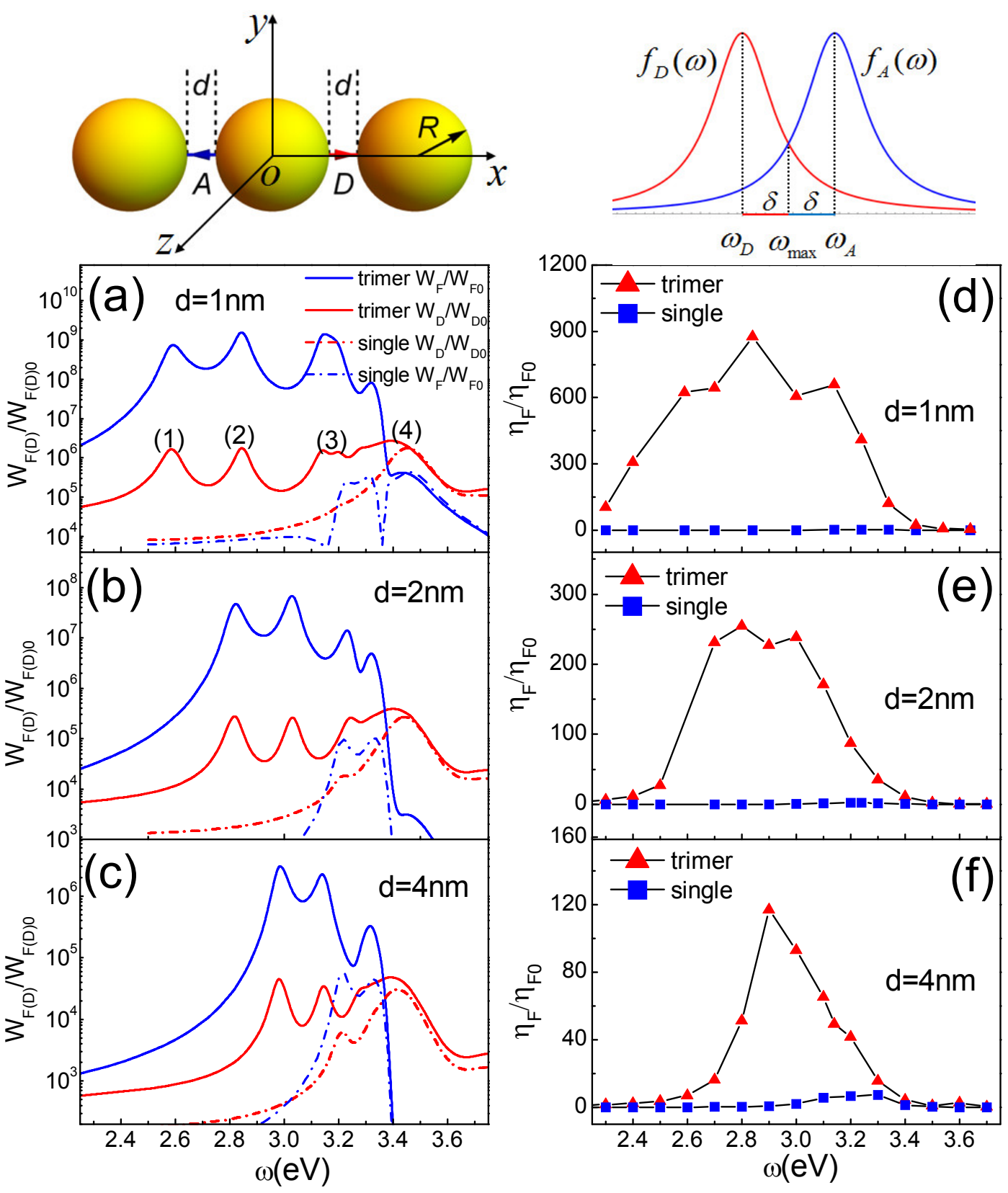

FIG. 1 (color online) FRET rate and efficiency of a pair of molecules located in the gaps of a nanosphere trimer with different separation distances (geometry and coordinate of the system are shown on left top of the figure). The radii of Ag spheres are taken as R=10nm. The left column corresponds to the FRET rate and the right column to the FRET efficiency. The blue and red solid lines in the left column represent $W_F / W_{F0}$ and $W_D / W_{D0}$, respectively. The corresponding results for the single sphere case (removing two spheres and keep the middle one) are described by dashed lines. The red triangle and blue square in the right column represent $\eta_F / \eta_{F0}$ as a function of maximum overlap frequency for the trimer and the single sphere case, respectively. (a) and (d) correspond to the case with d=1nm, (b) and (e) to that with d=2nm, (c) and (f) to that with d=4nm. The overlap of donor emission and acceptor absorption spectral is shown on right top of the figure.

The blue and red solid lines in Fig. 1(a)-(c) correspond to $W_F / W_{F0}$ and $W_D / W_{D0}$, respectively. For comparison, the corresponding results for the single sphere (removing the two spheres on both sides and keep the middle one in the trimer) are also shown as blue and red dashed lines. We can see clearly that $W_F / W_{F0}$ and $W_D / W_{D0}$ strongly depend on the frequency. For the case of the single sphere, the FRET rate (blue dashed line) can be improved by 4~5 orders, which is in agreement with the previous investigation.[44] At the same time, the donor decay rate has also been improved by the same order around the single scattering resonance frequency (3.45eV in such a case). As a result, the FRET efficiency is not improved as shown by blue squares in Fig. 1(d)-(f).

However, the situation becomes different for the trimer. Take the case with d=1nm as an example, the enhancement of the donor decay rate $W_D / W_{D0}$ possesses four resonance peaks marked by (1)-(4) in the red solid line of Fig. 1(a). These peaks originate from two kinds of plasmon resonance, single scattering resonance and coupling resonance. The peak (4) comes from the single scattering localized surface plasmon resonance, which is determined by the property of the single sphere and is not sensitive to the gaps. The other three peaks (peak (1), (2) and (3)) are caused by the coupling plasmon resonances, which are very sensitive to the separation distances between two spheres. With the increase of gaps, the number of peaks decreases.

Around the single scattering resonance frequency (3.45eV for the peak (4)), the $W_D / W_{D0}$ is always bigger than $W_F / W_{F0}$. From Eqs. (5), (7) and (11), we find that the FRET efficiency is not improved although the $W_F / W_{F0}$ has been improved by several orders, which is similar to the case of the single sphere. In contrast, around the coupling plasmon resonance frequency, the FRET rate (corresponding to the $W_F / W_{F0}$) is improved by 8~9 orders, which exceeds the enhancement of the donor decay ($W_D / W_{D0}$) about 3 orders. As a result, the FRET efficiency has been enhanced by hundreds of times as shown by red triangle dots in Fig. 1(d).

Such a phenomenon depends on the separation distance d. With the increase of d, both the improvement of donor decay rate and FRET rate decreases. In addition $W_F / W_{F0}$ decreases faster than $W_D / W_{D0}$, thus, the efficiency also decreases. For example, at d=1nm, the maximum of $\eta_F / \eta_{F0}$ is close to 900, however, it becomes 120 at d=4nm. When the separation distances become very large, the coupling resonance peaks disappear and the phenomenon degenerates to the case of the single sphere.

The above results only focus on a kind of the spectral overlap and orientations of dipole moments of molecules. In fact, the spectral overlap and orientations of dipole moments of molecules have important effects on the absolute FRET rate in vacuum[9], however, in the present work we focus on the enhancement of the FRET rate and efficiency, therefore, the results described above always hold even though the spectral overlap and orientations change. The above results indicate that we can realize simultaneously giant enhancement of the FRET rate and efficiency in a wide wavelength range by using the nanosphere trimer with suitable parameters.

In order to further disclose the physical origin of the above phenomena, in Fig. 2 we plot the static electric field patterns of an excited donor molecule with unit electric dipole moment located in one gap of the nanosphere trimer under different resonance frequencies. Detailed calculation process can be found in supplementary material. The coordinate of the system has been shown on the left top of Fig. 1. Fig. 2(a)-(c) display the results with d=1nm at resonance frequencies $\omega$=2.84eV, 3.15eV and 3.45eV, respectively, Fig. 2(d)-(f) correspond to the case with d=4nm at $\omega$= 3.0eV, 3.14eV and 3.45eV. Fig. 2(c) and (f) describe the case of the single scattering resonance, the others correspond to the coupling resonances.

It is seen clearly that at the single scattering resonance the field in the acceptor position is almost zero, although it is very strong around the position. This is in contrast to the cases described in Fig. 2(a) and (b). In such a case, the field intensity in the position of the acceptor is very close to the value at the donor position, that is to say, the acceptor can strongly feel the donor through the coupling resonance of the nanosphere cluster. Such a phenomenon is very sensitive to the gap. With the increase of the gap, the field in the position of the acceptor decrease rapidly. For example, the field intensities at positions of acceptors in Fig. 2(d) and (e) are ten times smaller than those in Fig. 2(a) and (b), which corresponds to the values of the FRET rate and efficiency in two cases. The above discussions only focus on the resonance cases. For the non-resonance cases, our calculated results show that the fields in the positions of both donor and acceptor are very weak, the FRET rate and efficiency have become relatively low. Take d=4nm as an example, from Fig. 1(c), it is seen clearly that the resonance range is about 2.8eV~3.5eV. When the frequency is less than 2.8eV or more than 3.5eV, the enhancements of FRET rate and donor decay rate drop sharply, which leads to the remarkable decrease of the FRET efficiency. For instance, $\eta_F / \eta_{F0}$ is near zero at $\omega$=2.4eV. These results clearly show that strong coupling resonances are the origin of the simultaneously large enhancement for the FRET rate and efficiency.

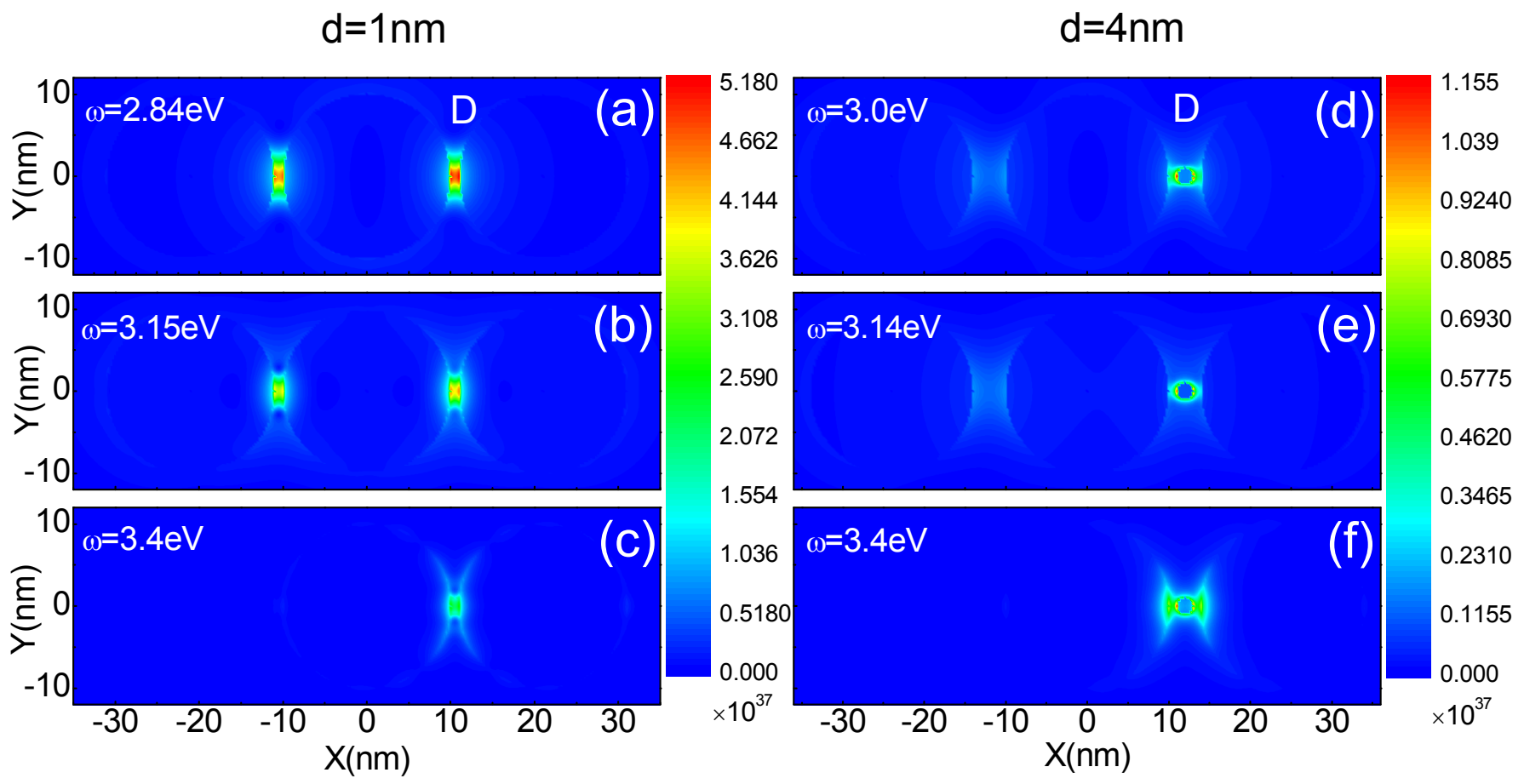

FIG. 2 (color online) Field patterns of an excited donor molecule with unit electric dipole moment located in the gap of the nanosphere trimer under different resonance frequencies. The left column corresponds to the case with d=1nm and the right column to that with d=4nm. (a), (b), (d) and (e) describe the cases at the coupling resonance frequencies $\omega = 2.84eV$, $3.15eV$, $3.0eV$ and $3.14eV$, respectively. The corresponding results at the single scattering resonance frequency $\omega = 3.45eV$ are given in (c) and (f). The other parameters are taken identical with those in Fig. 1.

## IV.THE EFFECT OF NANOPARTICLE SIZES AND NON-LOCAL RESPONSE

The above discussions only focus on the case with R=10nm. In fact, the FRET rate and efficiency also depend on the sizes of spheres when the separation distances are fixed. Recent investigations have shown that as the nanoparticle radii goes below 10 nm or the gap between two metal spheres is smaller than 1.0nm, the effect of quantum size or nonlocal response on the plasmon resonance becomes important [49-52]. In this part, we employ a hydrodynamic model to study the effect of nonlocal response on the energy transfer. For the single sphere, the field distribution could be solved exactly according to Ref. [49]. For the trimer system in this paper, the nonlocal effect has been included exactly for every sphere and the total effect has been considered by using the multi-scattering method as shown in supplementary material. The Fermi velocity for the metal Ag is taken as $V_F = 1.39\times10^6 m/s$, and the transverse response is $\varepsilon(\omega) = \varepsilon_\infty + \sigma(\omega)/i\varepsilon_0\omega$, where bound response $\varepsilon_\infty = 1.0$, Drude conductivity $\sigma(\omega) = i\varepsilon_0\omega_P^2/(\omega + i\eta)$ with plasma frequency $\omega_P = 9.01eV$ and loss rate $\eta = 0.048eV$ [49].

Figure 3(a)-(c) display $W_F/W_{F0}$ for Ag trimers with R=5nm, 10nm and 15nm at d=1nm, respectively, the corresponding results for $\eta_F/\eta_{F0}$ as a function of maximum overlap frequency are given in Fig. 3(d)-(f). The black lines represent the calculated results without the nonlocal effect and the red lines are those with the nonlocal effect. Comparing them, we find that considering the nonlocal effect resonance peaks for $W_F / W_{F0}$ and $\eta_F / \eta_{F0}$ not only show blue shift, their amplitudes also decrease. However, the phenomena for simultaneously giant enhancement of the FRET rate and efficiency always appear. In addition, the spectral width at the high frequency to appear the above phenomena can be extended due to the nonlocal bulk plasmon excitation. With the increase of radii of spheres at fixed gaps, the nonlocal effect on the phenomenon becomes smaller and smaller, the enhancement of the FRET rate and efficiency by coupling resonance plasmonic excitations is significantly increased. For example, there is an order of $\eta_F / \eta_{F0}$ improvement for the case with R=15nm compared with that of R=10nm.

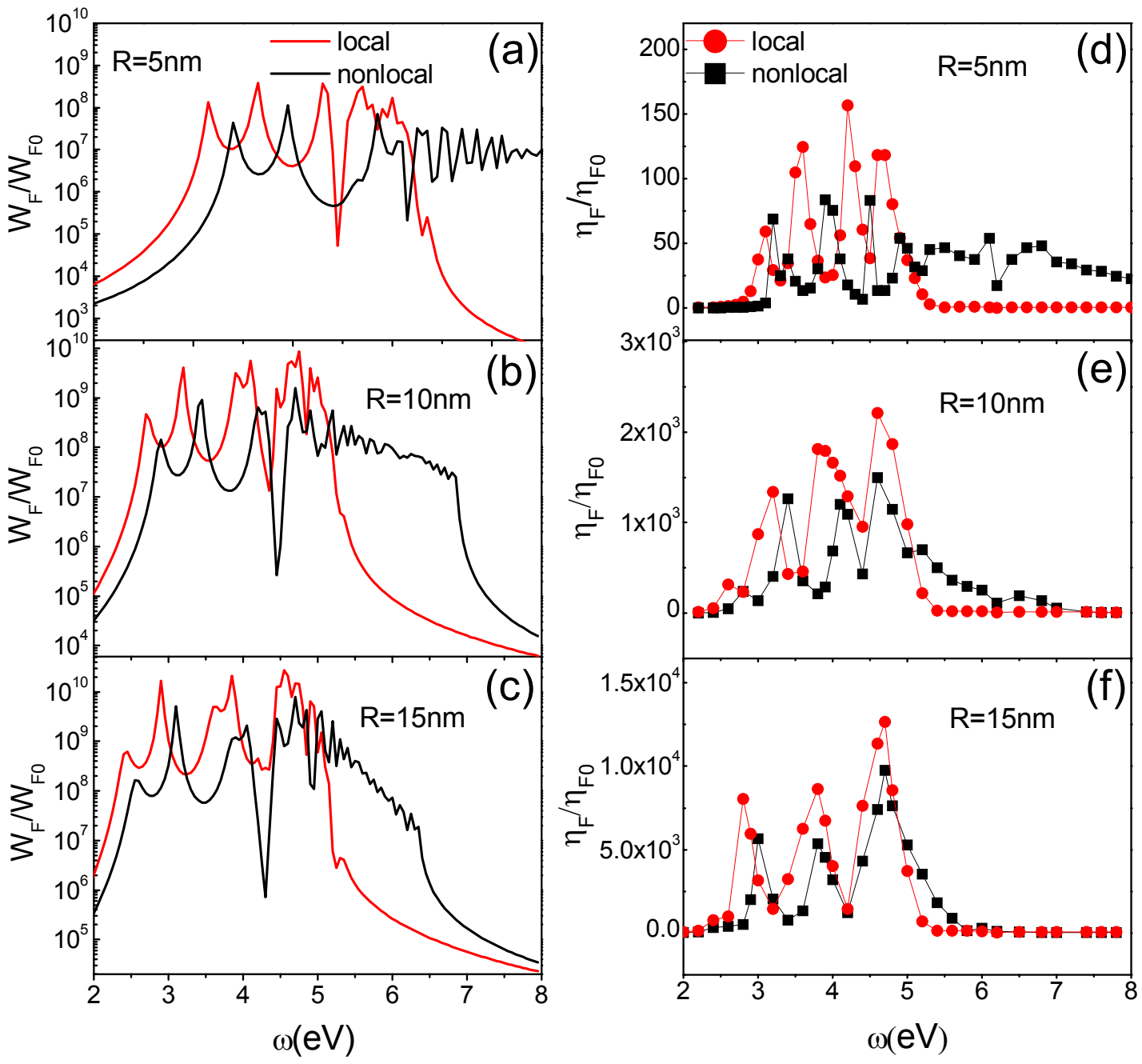

FIG. 3 (color online) FRET rate and efficiency of a pair of molecules located in the gaps of a nanosphere trimer with various sizes of spheres. (a), (b) and (c) represent $W_F / W_{F0}$ for the Ag trimer with R=5nm, 10nm and 15nm at d=1nm, respectively. (d), (e) and (f) describe the corresponding results for $\eta_F / \eta_{F0}$ as a function of maximum overlap frequency. The black lines correspond to the results without the nonlocal effect, the red lines to those with the nonlocal effect.

## V. THE RELATION BETWEEN THE LDOS AND FRET RATE

The question of whether the FRET rate depends on the optical LDOS remains under debate. In the following, we study the relation between the FRET rate and the optical LDOS in the above system. It is well known that the spontaneous decay rate is proportional to the imaginary part of the Green's tensor of the medium at the molecular position, which has a linear relation with the optical LDOS [53]. Thus, calculating the decay rate $K_D / K_{D0}$ is equivalent to the calculation of the optical LDOS. We plot the calculated results for the normalized FRET rate $K_F / K_{F0}$ and efficiency $\eta_F / \eta_{F0}$ as a function of the normalized donor decay rate $K_D / K_{D0}$ in Fig. 4(a) and (b), respectively. Here the parameters are taken identical with those in Fig.1. The change of the donor decay rate can be realized by altering the separation distance d, the dependences of the FRET rate and efficiency on the half of the separation distance d/2 are given in Fig. 4 (c) and (d). The blue square and red triangle represent the cases for the coupling resonance frequency ( $\omega_{\max} = 3.15eV$ ) and single scattering frequency ( $\omega_{\max} = 3.45eV$ ), respectively. For comparison, the single sphere case with the resonance frequency $\omega_{\max} = 3.45eV$ is also given as black circle lines.

We can see clearly that the energy transfer rate is almost independent of the LDOS at the single scattering resonance for the trimer or the single sphere case at the resonance frequency, which is similar to some previous investigations.[27,28,37] In such a case, the efficiency has no apparent enhancement with the increase of the optical LDOS. In contrast, it exhibits rich phenomena for the coupling resonance frequencies in the nanoparticle cluster. When gaps are relatively large, the FRET rate increases linearly with the increase of the optical LDOS (see inset in Fig. 4(a)). When gaps are very small, like d=1nm or 2nm, the FRET rate and efficiency under the coupling resonance frequency have exponential relationships with the LDOS (Fig. 4(a)). In such a case, the efficiency also increases exponentially with the increase of the optical LDOS (Fig. 4(b)). The origin of these phenomena is also attributed to the coupling plasmonic resonances in the nanosphere cluster as has been disclosed in Fig. 2.

Our calculated results indicate that the dependence of the energy transfer rate on the LDOS is determined by the system and condition. Based on the analysis above, independent, linear and exponential relations have been observed in the trimer systems under various system parameters. We believe that our findings are helpful in understanding the debated question. Recently, some clusters of gold nanospheres, i.e. the dimers/trimers/tetramers, were fabricated successfully by using cysteine chiral molecules as linkers at the hotspots[54]. We expect the phenomena in this work could be confirmed experimentally in the future. In the end, we would like to point out that, although the phenomena disclosed in the present paper are for the specific trimer system, they are also applicable to other cases if the similar

"coupled resonance" can be realized in these cases.

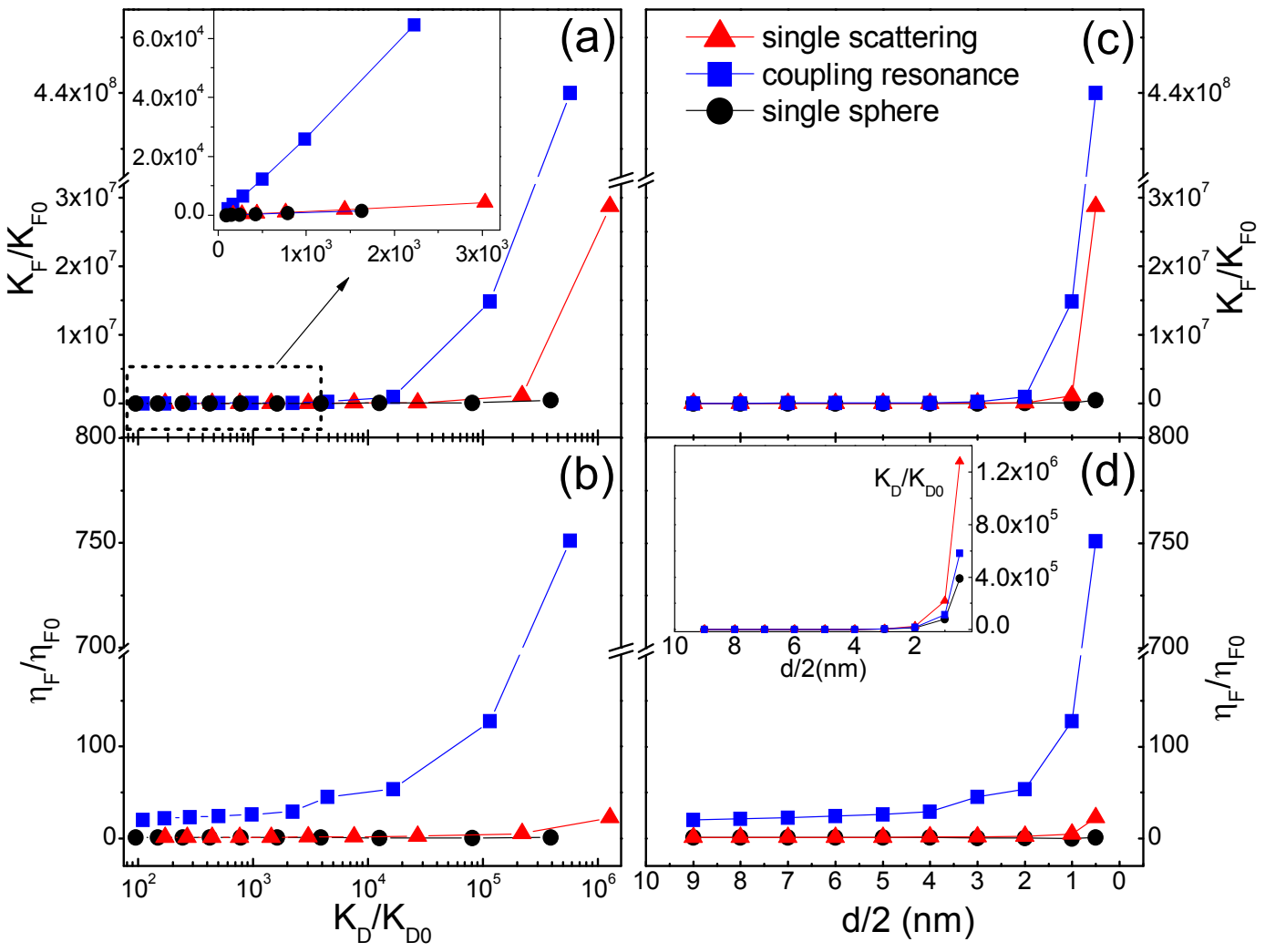

FIG. 4 (color online) The normalized FRET rate $K_F / K_{F0}$ (a) and efficiency $\eta_F / \eta_{F0}$ (b) as a function of the normalized donor decay rate $K_D / K_{D0}$. Blue squares represent the calculated results for the coupling resonance frequency at $\omega_{\max} = 3.15eV$, red triangles are the results for the single scattering resonant frequency at $\omega_{\max} = 3.45eV$ and black circles describe the single sphere case at $\omega_{\max} = 3.45eV$. (c) and (d) display the corresponding $K_F / K_{F0}$ and $\eta_F / \eta_{F0}$ as a function of molecule-sphere distance d/2, respectively. The other parameters are identical with those in Fig.1.

## VI. SUMMARY

The FRET rate and efficiency of an excited donor and a ground-state acceptor located in the gaps of a nanoparticle cluster have been studied by using a rigorous first-principles electromagnetic Green's tensor technique. Nonlocal effect has been considered by using the hydrodynamic model. It has been found that the FRET rate and efficiency can be enhanced simultaneously by more than 9 and 3 orders of magnitude at suitable parameters, respectively. The phenomenon originated in the coupling plasmonic resonances. This provides a new way to realize the energy transfer process with simultaneous ultrafast and ultrahigh efficiency. Furthermore, various dependences of the FRET rate on the LDOS have been observed in the same system under different conditions. At the coupling resonances, the dependence of the FRET rate is changed from linear to exponential. At the single scattering resonance, the energy transfer rate is almost independent of the LDOS. This means that some unresolved debates on how the FRET rate depends on the LDOS can be understood in a unified manner.

## ACKNOWLEDGMENTS

This work was supported by the National Key Basic Research Special Foundation of China under Grant 2013CB632704 and the National Natural Science Foundation of China (11574031 and 61421001).

## RERERENCES

[1] S. Angers, A. Salahpour, E. Joly, S. Hilairet, D. Chelsky, M. Dennis, and M. Bouvier, Proceedings of the National Academy of Sciences **97**, 3684 (2000).

[2] R. M. Clegg, in *Laboratory Techniques in Biochemistry and Molecular Biology* (Elsevier, 2009), pp. 1.

[3] A. Miyawaki, J. Llopis, R. Heim, J. M. McCaffery, J. A. Adams, M. Ikura, and R. Y. Tsien, Nature **388**, 882 (1997).

[4] D. M. Willard, T. Mutschler, M. Yu, J. Jung, and A. Orden, Anal. Bioanal. Chem. **384**, 564 (2006).

[5] S. John and J. Wang, Phys. Rev. B **43**, 12772 (1991).

[6] B. W. Lovett, J. H. Reina, A. Nazir, and G. A. D. Briggs, Phys. Rev. B **68**, 205319 (2003).

[7] T. Unold, K. Mueller, C. Lienau, T. Elsaesser, and A. D. Wieck, Phys. Rev. Lett. **94**, 137404 (2005).

[8] D. L. Dexter, J. Chem. Phys. **21**, 836 (1953).

[9] T. Förster, Ann. der Phys. **437**, 55 (1948).

[10] D. L. Andrews, Chem. Phys. **135**, 195 (1989).

[11] R. van Grondelle, J. P. Dekker, T. Gillbro, and V. Sundstrom, Biochimica et Biophysica Acta (BBA) - Bioenergetics **1187**, 1 (1994).

[12] R. Hildner, D. Brinks, J. B. Nieder, R. J. Cogdell, and N. F. van Hulst, Science **340**, 1448 (2013).

[13] W. Khlbrandt and D. N. Wang, Nature **350**, 130 (1991).

[14] D. K. Chatterjee, A. J. Rufaihah, and Y. Zhang, Biomaterials **29**, 937 (2008).

[15] J. Zhou, Z. Liu, and F. Li, Chem. Soc. rev. **41**, 1323 (2012).

[16] M. A. Baldo, M. E. Thompson, and S. R. Forrest, Nature **403**, 750 (2000).

[17] V. Vohra, G. Calzaferri, S. Destri, M. Pasini, W. Porzio, and C. Botta, ACS Nano **4**, 1409 (2010).

[18] S. Chanyawadee, R. T. Harley, M. Henini, D. V. Talapin, and P. G. Lagoudakis, Phys. Rev. Lett. **102**, 077402 (2009).

[19] D. J. Farrell and N. J. Ekins-Daukes, Nat Photon **3**, 373 (2009).

[20] K. Shankar, X. Feng, and C. A. Grimes, ACS Nano **3**, 788 (2009).

[21] Z. Yang, X. Zhou, X. Huang, J. Zhou, G. Yang, Q. Xie, L. Sun, and B. Li, Optics Lett. **33**, 1963 (2008).

[22] A. Konrad, M. Metzger, A. M. Kern, M. Brecht, and A. J. Meixner, Nanoscale **7**, 10204 (2015).

[23] F. Schleifenbaum, A. M. Kern, A. Konrad, and A. J. Meixner, Phys. Chem. Chem. Phys. **16**, 12812 (2014).

[24] K. H. An, M. Shtein, and K. P. Pipe, Optics Express **18**, 4041 (2010).

[25] P. Andrew and W. L. Barnes, Science **290**, 785 (2000).

[26] D. Bouchet, D. Cao, R. Carminati, Y. De Wilde, and V. Krachmalnicoff, Phys. Rev. Lett. **116**, 037401 (2016).

[27] C. Blum, N. Zijlstra, A. Lagendijk, M. Wubs, A. P. Mosk, V. Subramaniam, and W. L. Vos, Phys. Rev. Lett. **109**, 203601 (2012).

[28] M. J. A. de Dood, J. Knoester, A. Tip, and A. Polman, Phys. Rev. B **71**, 115102 (2005).

[29] P. Ghenuche, J. de Torres, S. B. Moparthi, V. Grigoriev, and J. Wenger, Nano Lett. **14**, 4707 (2014).

[30] P. Ghenuche, M. Mivelle, J. de Torres, S. B. Moparthi, H. Rigneault, N. F. Van Hulst, M. F. García-Parajó, and J. Wenger, Nano Lett. **15**, 6193 (2015).

[31] V. K. Komarala, A. L. Bradley, Y. P. Rakovich, S. J. Byrne, Y. K. Gun'ko, and A. L. Rogach, Appl. Phys. Lett. **93**, 123102 (2008).

[32] M. Lunz, V. A. Gerard, Y. K. Gun'ko, V. Lesnyak, N. Gaponik, A. S. Susha, A. L. Rogach, and A. L. Bradley, Nano Lett. **11**, 3341 (2011).

[33] M. Lunz, X. Zhang, V. A. Gerard, Y. K. Gun'ko, V. Lesnyak, N. Gaponik, A. S. Susha, A. L. Rogach, and A. L. Bradley, J. Phys. Chem. C **116**, 26529 (2012).

[34] D. Martín-Cano, L. Martín-Moreno, F. J. García-Vidal, and E. Moreno, Nano Lett. **10**, 3129 (2010).

[35] T. Nakamura, M. Fujii, K. Imakita, and S. Hayashi, Phys. Rev. B **72**, 235412 (2005).

[36] V. N. Pustovit and T. V. Shahbazyan, Phys. Rev. B **83**, 085427 (2011).

[37] F. T. Rabouw, S. A. den Hartog, T. Senden, and A. Meijerink, Nat Commun **5** (2014).

[38] F. Reil, U. Hohenester, J. R. Krenn, and A. Leitner, Nano Lett. **8**, 4128 (2008).

[39] R. G. West and S. M. Sadeghi, J. Phys. Chem. C **116**, 20496 (2012).

[40] Y.-C. Yu, J.-M. Liu, C.-J. Jin, and X.-H. Wang, Nanoscale Research Lett. **8**, 209 (2013).

[41] J. Zhang, Y. Fu, and J. R. Lakowicz, J. Phys. Chem. C **111**, 50 (2007).

[42] X. Zhang *et al.*, ACS Nano **8**, 1273 (2014).

[43] L. Zhao, T. Ming, L. Shao, H. Chen, and J. Wang, J. Phys. Chem. C **116**, 8287 (2012).

[44] H. Y. Xie, H. Y. Chung, P. T. Leung, and D. P. Tsai, Phys. Rev. B **80**, 155448 (2009).

[45] T. Nakamura, M. Fujii, S. Miura, M. Inui, and S. Hayashi, Phys. Rev. B **74**, 045302 (2006).

[46] H. T. Dung, L. Knöll, and D.-G. Welsch, Phys. Rev. A **65**, 043813 (2002).

[47] J. A. Gonzaga-Galeana and J. R. Zurita-Sánchez, J. Chem. Phys. **139**, 244302 (2013).

[48] P. B. Johnson and R. W. Christy, Phys. Rev. B **6**, 4370 (1972).

[49] T. Christensen, W. Yan, S. Raza, A.-P. Jauho, N. A. Mortensen, and M. Wubs, ACS Nano **8**, 1745 (2014).

[50] N. A. Mortensen, S. Raza, M. Wubs, T. Søndergaard, and S. I. Bozhevolnyi, Nat Commun **5** (2014).

[51] J. A. Scholl, A. L. Koh, and J. A. Dionne, Nature **483**, 421 (2012).

[52] C. David and F. J. García de Abajo, J. Phys. Chem. C **115**, 19470 (2011).

[53] L. Novotny and B. Hecht., *Principles of Nano-Optics* (Cambridge University Press, 2012).

[54] R.-Y. Wang *et al.*, J. Phys. Chem. C **118**, 9690 (2014).